# Template Dissolution Interfacial Patterning of Single Colloids for Nanoelectrochemistry and Nanosensing

Joong Bum Lee[1,†], Harriet Walker[2,†], Yi Li[3], Tae Won Nam[1], Aliaksandra Rakovich[4], Riccardo Sapienza[2], Yeon Sik Jung[1], Yoon Sung Nam[1,5*], Stefan A. Maier[2,6], and Emiliano Cortés[6*]

[1]Department of Materials Science and Engineering, Korea Advanced Institute of Science and Technology, Daejeon 34141, Republic of Korea

[2]The Blackett Laboratory, Department of Physics, Imperial College London, London SW7 2AZ, United Kingdom

[3]School of Microelectronics, MOE Engineering Research Center of Integrated Circuits for Next Generation Communications, Southern University of Science and Technology, Shenzhen, 518055 Guangdong, China

[4]Department of Physics, King's College London, London WC2R 2LS, United Kingdom

[5]KAIST Institute for Nanocentury, Korea Advanced Institute of Science and Technology, Daejeon 34141, Republic of Korea

[6]Faculty of Physics, Ludwig-Maximilians-Universität München, 80539 München, Germany

[†] These authors contributed equally to this work

[*] Email: *yoonsung@kaist.ac.kr* and *emiliano.cortes@lmu.de*

ABSTRACT

Deterministic positioning and assembly of colloidal nanoparticles (NPs) onto substrates is a core requirement and a promising alternative to top down lithography to create functional nanostructures and nanodevices with intriguing optical, electrical, and catalytic features. Capillary-assisted particle assembly (CAPA) has emerged as an attractive technique to this end, as it allows controlled and selective assembly of a wide variety of NPs onto predefined topographical templates using capillary forces. One critical issue with CAPA, however, lies in its final printing step, where high printing yields are possible only with the use of an adhesive polymer film. To address this problem, we have developed a template dissolution interfacial patterning (TDIP) technique to assemble and print single colloidal AuNP arrays onto various dielectric and conductive substrates in the absence of any adhesion layer, with printing yields higher than 98%. The TDIP approach grants direct access to the interface between the AuNP and the target surface, enabling the use of colloidal AuNPs as building blocks for practical applications. The versatile applicability of TDIP is demonstrated by the creation of direct electrical junctions for electro- and photoelectrochemistry and nanoparticle-on-mirror geometries for single particle molecular sensing.

**Keywords**

nanoparticles, capillary-assisted particle assembly, localized surface plasmon resonance, nano-electrodes, charge transfer, SERS

Metal nanoparticles (NPs) are fundamental building blocks of nanotechnology with interesting optical, electrical, magnetic, and catalytic properties for a multitude of applications.[1-3] To design and engineer functional nanodevices that effectively harness these properties, it is imperative to assemble NPs onto substrates in a deterministic manner. Achieving this using colloidal NPs is highly desirable as they are superior to lithographically fabricated counterparts in terms of crystallinity, large scale production, and surface tunability.[4-6] However, an open challenge remains to develop techniques for controllable and accurate spatial arrangement of colloids both at the clustered and single particle level. To this end, a host of innovative methods have been proposed such as template-assisted self-assembly,[7-10] DNA-mediated assembly,[11,12] optical printing,[13,14] electrostatic assembly,[15,16] and electrophoretic deposition (EPD).[17,18] Although these existing techniques offer a broad prospect for colloidal patterning, many problems still exist such as long assembly times, limited material selection for particles and substrates, limited scalability, and post-assembly particle damage. In addition to these methods, capillary-assisted particle assembly (CAPA)[19,20] has emerged as a technique that combines the advantages of bottom-up synthesis of colloidal NPs and top-down nanofabrication of a topographical template created with high precision. In this technique, NPs are selectively deposited as per predefined topographical patterns and thus controllably assembled with respect to orientation and position. More specifically, CAPA involves the confinement of a colloidal NP suspension in between a topographical template and a unidirectional sliding plate that forms a receding meniscus, wherein capillary and convective forces act at the vapor/liquid solution/solid substrate contact line to confine the NPs into the patterned traps.[19] Efforts to understand the working principles of the method have uncovered that the assembly process is influenced by various parameters such as temperature, assembly speed, trap geometry, and particle concentration.[21,22] For metal nanostructures in particular, previous reports demonstrated the assembly of nanospheres,[19,20,23,24] nano-cuboctahedra,[25] nanorods,[26,27] and heterogeneous hybrid clusters[28] with the primary focus on optimizing the conditions to achieve high-yield assembly. With the assembly of single particles and multimeric clusters, these reports have shown that CAPA has no limit to how close individual particles can be positioned relative to one another as the particle traps can be designed accordingly. Some applications of CAPA assemblies include surface-enhanced Raman spectroscopy (SERS),[29] enhanced second harmonic generation (SHG),[30] and security features,[31] all demonstrated with NPs as-deposited inside the topographical templates. It should also be noted, however, that having NPs positioned inside predefined traps only allows partial access to the NPs, hindering their use for most applications.

To expand further the use of the CAPA method, it would be necessary to print the assembled NPs onto substrates in free-standing form, increasing their accessibility. In standard CAPA technique where polydimethylsiloxane (PDMS) is used to replicate the topographical Si master, the assembled nanoparticles are printed by bringing the PDMS stamp into direct contact with substrates such as Si or glass followed by stamp release. However, the low surface energy of PDMS and its poor adhesion to substrates[32] leads to severely low yields in printing isolated, single nanoparticles unless a thin polymeric adhesion layer is applied to the receiving substrate.[20] The presence of an adhesion layer poses a significant limitation on the range of practical applications the technique can be translated to, where direct interaction between the NP and the substrate is required. For instance, having an adhesion layer can interfere with plasmonic properties by altering the local refractive index, electrical measurements by impeding direct charge transfer, and sensing of probe molecules ideally positioned at hot spots formed at the NP and substrate gap. To tackle this issue, we herein present template dissolution interfacial patterning (TDIP), a technique which allows colloidal NPs to be assembled and printed on various receiver substrates in free-standing form at the single particle level. In the TDIP technique, the topographical Si master is replicated using poly(methyl methacrylate) (PMMA) as an alternative to the PDMS stamp used in nearly every study concerning CAPA. The PMMA replica used in this work is derived from a dilute polymer solution and created as thin film that facilitates the NPs to come easily in direct contact with surfaces. By simple dissolution of the PMMA replica in acetone and hence omitting the stamp releasing step required in standard CAPA technique, the NPs are printed with a very high yield and accuracy onto receiver substrates. The applicability of TDIP, which grants direct particle-substrate interactions, is demonstrated by creating a nano-electrode for electro- and photoelectrochemistry as well as a nanoparticle-on-mirror construct for single particle surface-enhanced Raman spectroscopy (SERS). These results present the possibility to implement TDIP for sensing, nano-electrochemistry, opto-electronics and metamaterials, among others, utilizing single colloidal nanoparticles as building blocks.

**RESULTS AND DISCUSSION**

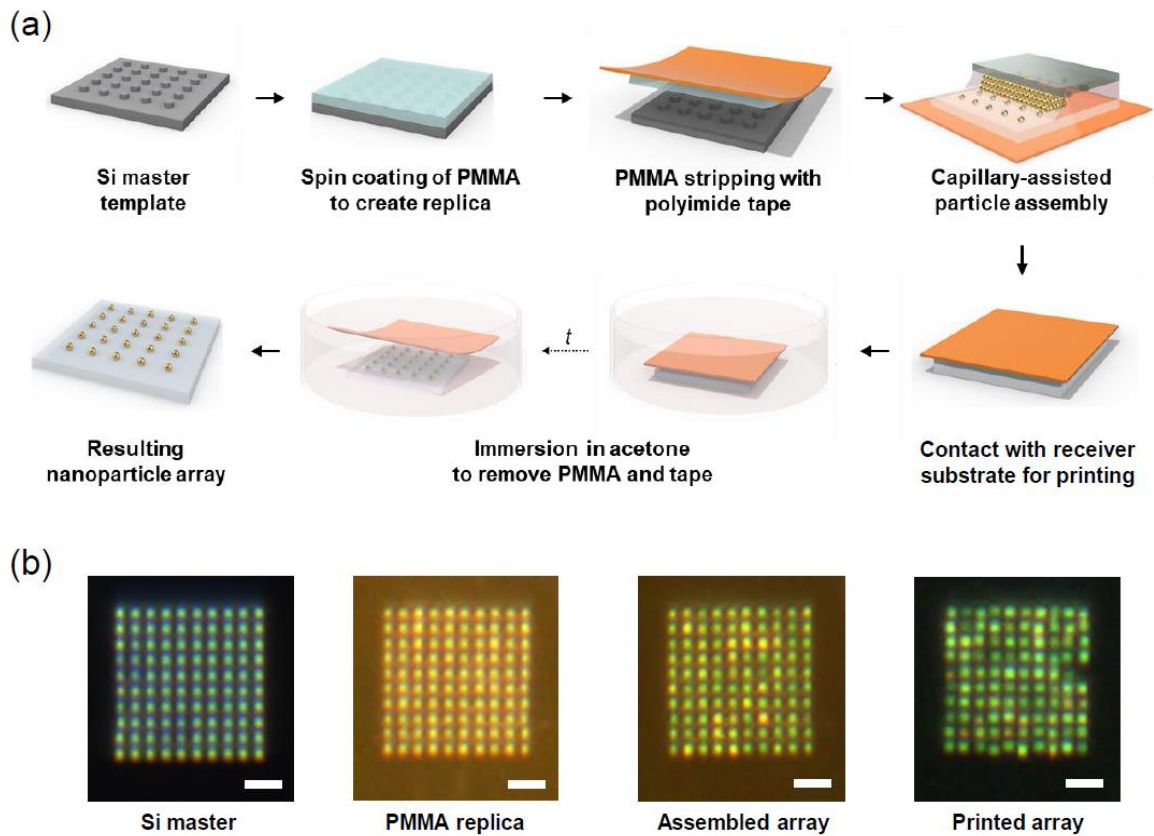

**Figure 1. Overall process of the template dissolution interfacial patterning (TDIP) technique.** (a) Schematic illustration of the assembly and printing of single AuNPs. (b) Dark-field images of the master, PMMA replica, AuNPs assembled on PMMA, and printed AuNP array. Scale bars: 2 µm.

Figure 1(a) shows the schematic illustration of the TDIP process, using PMMA as the assembly template. The Si master with patterned arrays of nano-pillars was initially treated with hydroxy-terminated PDMS brush whose low surface tension allows the spin-coated PMMA film to be easily and cleanly stripped using polyimide (PI) tape.[33] Once the PMMA replica is prepared as such, the Si master can be reused indefinite number of times to create the same replica template (*e.g.*, used more than 200 times for this work). Next, CAPA was used to assemble 100 nm diameter AuNPs into the wells of the PMMA template. The underlying condition for successful assembly using CAPA is to create a high density accumulation zone (AZ) at the edge of the meniscus.[34] The formation of AZ hinders Brownian motion and diffusivity of the NPs in solution, thereby ensuring that particles remain trapped at the desired site for deposition.[21] To print the assembled NPs we use immersion transfer printing.[35] This printing technique is based on an adhesion switching mechanism that considers the interaction energies at the interface between each individual component involved in the assembly and transfer of nanoparticles. First, the PI tape is manually pressed onto a receiver substrate to bring the nanoparticles in direct contact with the substrate surface. The substrate is then immersed entirely in acetone, which acts to dissolve the PMMA over time by polymer chain disentanglement, owing to their similar solubility parameter values.[36] This weakens the pre-existing force between the NP and the PMMA template, allowing weak van der Waals forces between the NP and the substrate to dominate. As the PMMA is removed by dissolution in acetone the PI tape is naturally detached from the surface, hence leaving only the NPs adsorbed on the substrate. Considering the expression for van der Waals interaction energy ($V_A$) between a spherical particle of radius $R$ near a flat surface,

$$V_A = -\frac{AR}{6H}$$

where $A$ is the Hamaker constant and $H$ is particle-substrate distance, the $V_A$ between the NP and the receiver substrate must be greater than the solubility energy of citrate-capped AuNP-acetone to remain adsorbed in free-standing form on the solid surface. Details of the main steps of the entire assembly and printing process are provided in Figure 1(b), which include the dark-field images of the master template, PMMA replica, and examples of assembled and printed AuNP arrays. Here, for consistency with imaging non-transparent substrates (*e.g.*, Si master), a reflected dark-field configuration with an air objective (100X, 0.9 NA) was used to collect scattered light from patterned arrays in each step.

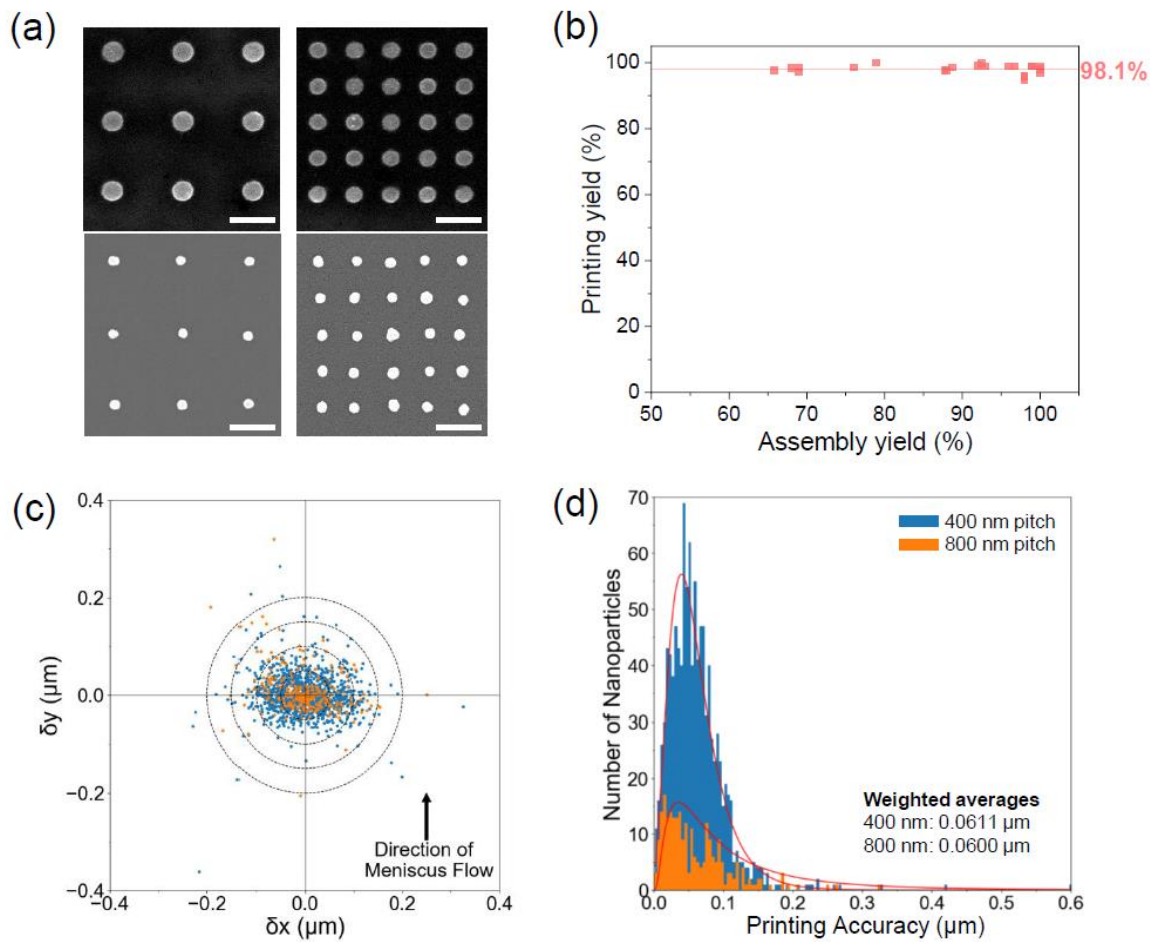

**Figure 2. Assembly and printing analysis of AuNP arrays printed on glass.** (a) SEM images of the Si master of both 800 nm and 400 nm pitches (top row) and corresponding AuNP arrays printed on glass (bottom row). Scale bars: 500 nm. (b) Printing yield of arrays plotted against assembly yield. (c) Bullseye plot of the *x-y* offset for each NP analyzed. (d) Histogram showing the printing accuracy for NPs assembled and printed on templates with 400 nm and 800 nm pitch.

So far, the printing step after NP assembly has not been scrutinized further due to the low printing yield in the absence of adhesion layers (*e.g.*, PMMA[20,31] or polyisobutylene[37]) or wetting of chemically-modified films.[38,39] To further elucidate the capabilities of this adhesive-free printing technique, single 100 nm AuNP arrays of both 800 nm and 400 nm pitch were assembled from a precisely defined master template, and both the assembly and printing yields were quantified. Figure 2(a) shows SEM images of both the printed AuNP arrays and the Si pillar arrays that were used to generate the printing template. To keep the NPs deposited as an array on PMMA pristine for the printing process, SEM imaging after assembly was omitted because doing so would require coating with a thin conductive layer. Instead, either bright-field or dark-field images were taken after the assembly to first determine the assembly yield and to use later as a reference to compare with the SEM of printed arrays to quantify the printing yield. An example of how the assembly and printing yields were calculated are provided in the Figure S1. Using this method, the yields for 22 different arrays (total of 3450 particles) were plotted in a scatter graph, as given in Figure 2(b). The figure shows that regardless of the assembly yield, which ranges from above 68% and is subject to improvement for nanoscale CAPA,[28] the printing yield is kept consistently high at an average of 98.1%. The given printing yield is very competitive when compared to values in other studies where adhesion layers were used, typically reported to be over 95%.[20]

Next, to delve deeper into the final printing step statistically, the printing accuracy was determined. This was done by measuring the offset distances of the center of each individual NP in an array from the ideal position defined by the master template and its respective PMMA replica. By combining these into a single plot, a bullseye chart as in Figure 2(c) can be obtained in which the blue and orange points pertain to NPs from 400 nm pitch and 800 nm pitch, respectively. Figure S2 illustrates an example of the printing accuracy analysis, from the original SEM of the printed array to the final bullseye chart. Representing the individual points in bullseye as function of the offset distances into a histogram (Figure 2(d)), the mean accuracy of the printing was calculated to be 60.8 nm, and 90% of the printed NPs fall within a 107 nm radius of the optimal printing position. The printing accuracy for this technique is therefore better than that of the conventional approach using PDMS.[20] The mean accuracy value is also half the size of the NPs printed in this case, and 30% of the size of the template hole. This method also yields a printing accuracy comparable to that which the complementary technique of optical printing yields when analyzed by a similar metric of printing accuracy.[14] The printing accuracy of the technique is not significantly affected by the pitch of the array or by the size of the template hole. However, that is not to say that optimization of the trap size and pitch could not improve the printing accuracy further, as previously reported.[15,26] It is noted that the printing accuracy of the NPs is higher in the direction of the flow of the meniscus (standard deviation (s.d.) 41.9 nm) in comparison to the direction normal to the flow (s.d. 61.6 nm). The difference can be seen in Figure S2(d), where the data points are more scattered along the x-axis, normal to the assembly direction. This is likely due to the trapping force acting predominantly in the direction of the bulk meniscus flow, causing a break in symmetry in the printing accuracy parallel and normal to the meniscus flow.[21] The printing accuracy can potentially be lowered by establishing a more reproducible and controllable printing method such as incorporating a mechanical device capable of nanoscale alignment, rather than simple manual pressure. This would enable a more uniform pressure upon contact between the assembly template and the receiver substrate and help minimize the offsets. The statistical analysis on the printing step also highlights the importance of using a precisely defined master template for maximum fidelity to the desired pattern, which in this case was made possible by electron beam lithography. Different top-down approaches, such as photolithography, nanoimprint lithography or bio-inspired materials, could be used in the future to prepare the master template, adding greater flexibility to the current technique. Overall, the provided statistical analysis supports the high performance of the template dissolution method for accurate spatial arrangement of NPs. TDIP can be a versatile tool for patterning a large assortment of colloids in terms of size and shape. Provided the previous demonstration in printing 3-6 nm diameter quantum dots (albeit a different assembly approach),[35] we expect the technique to be feasible for sub-10 nm particles if assembly templates can be designed accordingly. In terms of particle geometry, TDIP can be extended to any non-spherical NPs obtainable through wet chemical synthesis and also potentially to non-metallic, micron-scale colloids.[40-43] As an example for non-spherical NPs, Figure S3 shows an array of printed nanocubes. Although the nanocube suspension used here has a relatively large polydispersity (Figure S3(a)) and contains other shapes (*e.g.*, triangles, rods, deformed nanocubes), patterning nanocubes is highly feasible considering that the surface contact area with the receiver substrate is in principle larger for nanocubes than for nanospheres. For anisotropic particles such as nanorods, the assembly template can be engineered with the correct library of shapes and sizes not just to trap the particles but also to precisely control their orientation.[26] Aside from orientation control, such optimization of the template can enable improved selectivity considering the polydispersity of the nanoparticle suspension. Carefully designed traps can effectively filter out any undesired sizes or shapes of particles that exist in solution.[21,44] It should also be noted that for the same purpose of having free standing arrays, an alternative method was considered where a thin PDMS mask with open holes is bonded to a glass substrate for NP deposition *via* CAPA, omitting the printing process (Figure S4 and Supplementary Note 1). Considering the two strategies in parallel, it is apparent that our current TDIP method to implement the printing step is much more effective in terms of assembly yield, time consumption, and precise positioning of NPs.

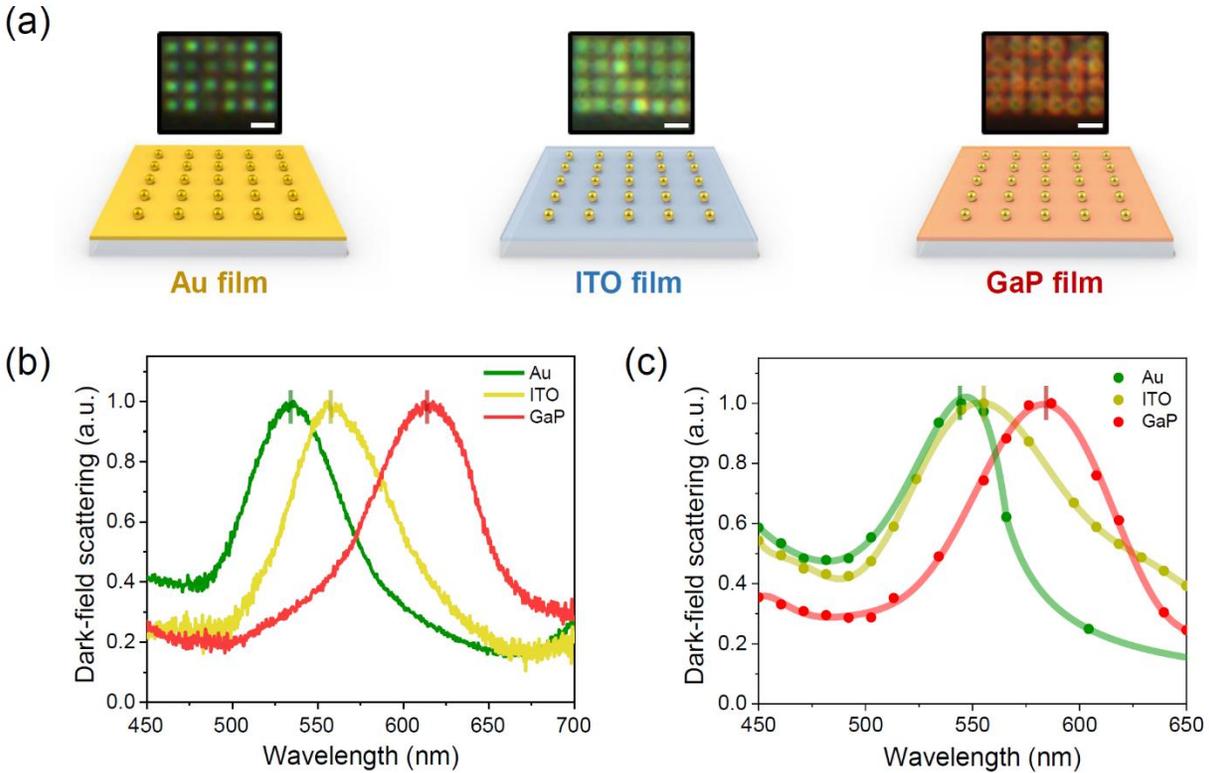

**Figure 3. Dark-field scattering response of AuNP arrays printed on different substrates.** (a) DF images of single AuNPs printed on Au film, ITO, and GaP, with different *z*-position adjustments made to find the best focus for each sample. Scale bars: 1 μm. Measured (b) and simulated (c) DF scattering spectra of single NPs printed on the different substrates.

With the absence of adhesion layers for printing, NPs can now be in direct contact with the substrate of interest. This allows CAPA to be translated to a much broader range of applications. As shown next, the optical properties of NPs can be tuned by printing them onto various target surfaces. Figure 3(a) shows both the visual illustration and the corresponding dark-field images of 100 nm AuNPs printed on Au film, indium tin oxide (ITO), and gallium phosphide (GaP) film substrates. The colors observed from the same NPs markedly differ within the visible range due to the change in the dielectric environment, which is related to the extinction spectrum, E(λ), of a metal NP based on the following equation:

$$E(\lambda) = \frac{24\pi^2 N \, a^3 \varepsilon_{med}^{\frac{3}{2}}}{\lambda \ln(10)} \left[ \frac{\varepsilon_i(\lambda)}{\left(\varepsilon_{re}(\lambda) + \chi \varepsilon_{med}\right)^2 + \varepsilon_i(\lambda)^2} \right]$$

where $\varepsilon_{med}$ is the dielectric constant of the surrounding medium, and $\varepsilon_{re}$ and $\varepsilon_i$ are the real and imaginary parts of the metal dielectric function. Given the relation $\varepsilon_{med} = n_{med}^2$ for dielectric constant $\varepsilon$ and refractive index $n$, the increase in the refractive index of the substrates ($n_{Au} < n_{ITO} < n_{GaP}$) induces a redshift of the plasmon resonance to satisfy the Fröhlich condition $\varepsilon_{re} = -2\varepsilon_{med}$.[45] It should also be noted that the AuNPs appear differently on each surface; solid green spot on Au film, blurry yellow spot on ITO, and doughnut-shaped red spot on GaP. For the AuNP on Au film case, which exhibits two plasmon resonances in the visible and in the near-infrared[46] (only the former displayed in Figure 3), the resonance from the green spot is attributed to the hybridization of the quadrupolar and dipolar plasmons of the NP coupled to the film substrate.[47] Given that the scattered radiation from the NP is characterized by oscillating fields with both vertical (*p*-mode) and horizontal (*s*-mode) polarized modes relative to the substrate, the augmented feature into doughnut shapes indicates that for AuNP on GaP the *p*-mode is more dominant.[46,48] This is supported by simulations of separate *s*- and *p*-polarized illumination which show a stronger response to *p*-polarized light for the AuNP on GaP case (Figure S5). With respect to the colors observed, the dark-field scattering spectra of single NPs in Figure 3(b) clearly show the shift of the resonance peaks, which are centered at 535 nm, 558 nm, and 615 nm for Au film, ITO, and GaP film, respectively. The significance of this result is that printing NPs directly onto different substrates allows us to finely tailor the resonance depending on the application. This underlines the advantage of TDIP that grants direct access to the interface between the NP and the substrate of interest. In conventional techniques that require an adhesive polymer film for high-yield printing, a shift in resonance cannot be achieved since the refractive index of the substrate (the polymer) remains constant. The trend of the resonance shifts is in par with that of the finite-difference time-domain (FDTD) simulations shown in Figure 3(c), with the difference in the maxima between the experiment and simulation slightly higher for GaP due to incremental changes in the optical constants of GaP, which is an evaporated polycrystalline film. The details of the simulations are provided in the **Methods** section.

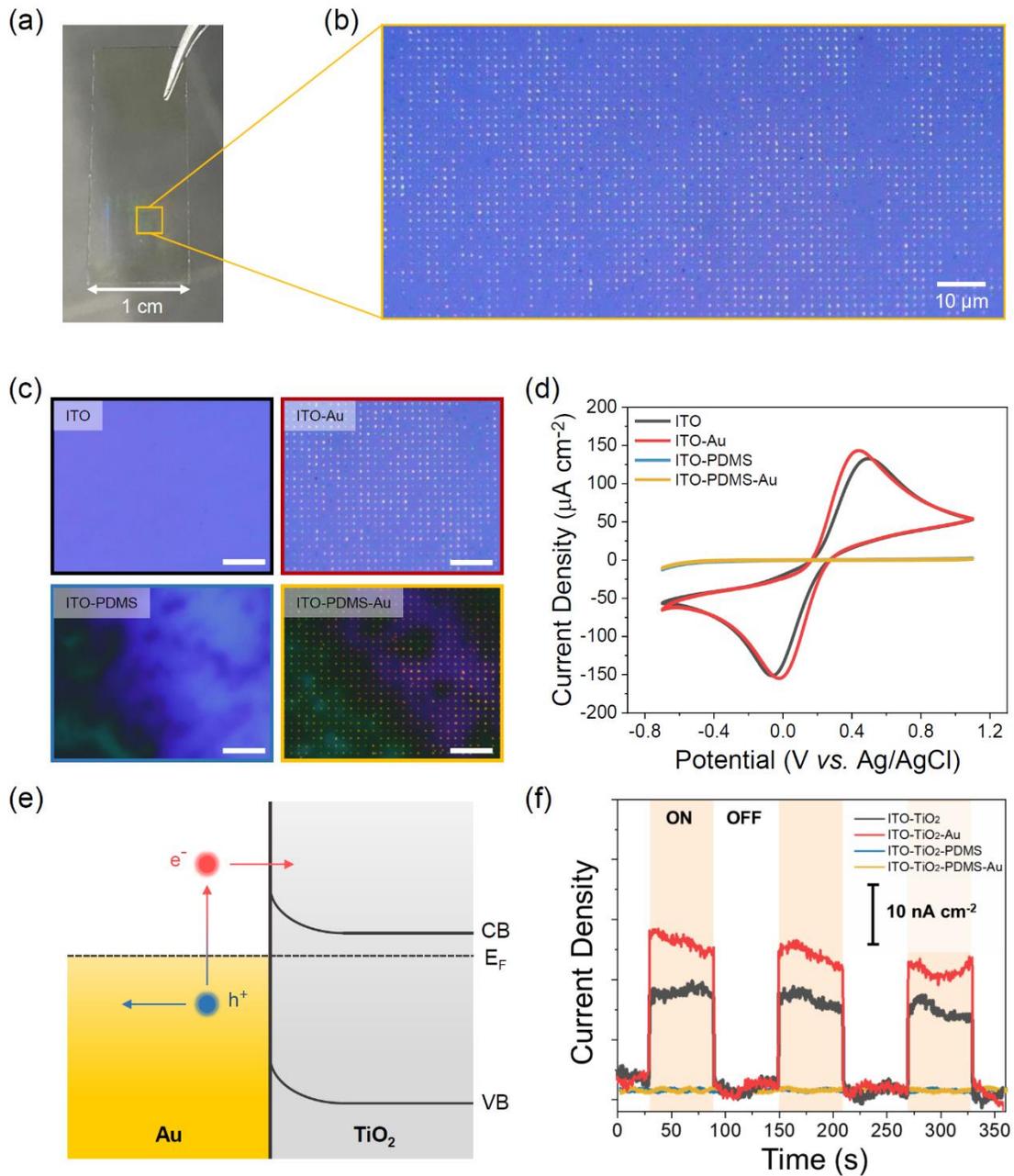

**Figure 4. Large area AuNP patterning for electro- and photoelectrochemistry.** (a) Photograph of large-scale AuNP assembly printed on ITO glass substrate. The structural colors from the AuNP pattern are visible to the naked eye. (b) Zoom-in optical image of a random area within the printed sample. (c) Bright-field optical images of ITO, ITO-Au, ITO-PDMS, and ITO-PDMS-Au. Scale

bars: 10 μm. (d) Cyclic voltammogram of four samples in (c) for ferrocyanide/ferricyanide redox couple, demonstrating the capability of technique to create direct electrical contact for charge transfer. (e) Schematic of band diagram of Au/TiO$_2$ Schottky junction, whose barrier can be overcome by plasmon-induced hot electrons in Au. (f) Photocurrent measurement in 0.1 M KH$_2$PO$_4$ containing 1 mM K$_3$Fe(CN)$_6$/K$_4$Fe(CN)$_6$ under visible light illumination.

As demonstrated in Figure 3, the capability to print AuNPs directly onto surfaces without an adhesion layer implies that a direct electrical contact can be created between the NP and conductive substrates such as ITO. With this, CAPA can be implemented for applications involving electrochemical processes, which could not be explored using conventional methods due to the presence of a polymeric adhesive film (*e.g.*, PMMA) for NP printing. As a simple demonstration to emphasize this advantage of the current technique, a nano-electrode was prepared for cyclic voltammetry (CV) analysis of a commonly used ferrocyanide/ferricyanide redox reaction using a three-electrode electrochemical cell, with Pt wire as the counter electrode and Ag/AgCl as the reference electrode. To create the nano-electrode, a large-scale array of AuNPs were printed on an ITO glass substrate (Figure 4(a)), and an optical image of a random area within the printed array is given in Figure 4(b). The CV profile of AuNP-printed ITO (labeled **ITO-Au**) was compared to those of bare ITO (labeled **ITO**), PDMS coated-ITO (labeled **ITO-PDMS**), and AuNPs printed on PDMS coated-ITO (labeled **ITO-PDMS-Au**). The optical images of the four different samples are provided in Figure 4(c). Here, the PDMS is a thin film spun cast on ITO that mimics the role of the adhesion layer which acts inevitably as an insulator, preventing electrical flow between the AuNP and ITO. The representative CV cycles for each electrode is plotted in Figure 4(d), with clear redox peaks only for ITO and AuNPs printed on ITO and none for PDMS-coated samples within the given potential window. By taking the difference between the anodic and cathodic peak potentials (peak-to-peak separation, $\Delta E_p$), the heterogenous electron transfer kinetics of the system can be deduced.[49] In this regard, the lower $\Delta E_p$ of **ITO-Au** (0.463 V) than that of **ITO** (0.560 V) also indicates that the electron transfer between the electrode and the electrolyte is faster when AuNP arrays decorate the ITO surface. The results herein indicate that the patterning of a nano-electrode is possible using AuNPs and feasible for electrochemical applications.

In the context of charge transfer capabilities, a natural progression in using AuNPs would be to utilize the hot electrons generated *via* the excitation of LSPR.[50] Thus, as a more direct quantitative analysis of charge transfer from AuNPs, photocurrent measurements were made based on a well-established $Au/TiO_2$ heterojunction for hot electron transfer systems.[51-53] Upon visible light excitation and subsequent plasmon damping, hot electron-hole pairs can be generated in AuNPs. As illustrated in Figure 4(e), the hot electrons have sufficient kinetic energy to be injected into the conduction band of $TiO_2$ by overcoming the Schottky barrier (typically around 1 eV[54]) at the $Au/TiO_2$ interface. Once assembled *via* CAPA, the AuNPs were printed on a 20 nm thick $TiO_2$ film deposited on ITO by electron beam evaporation. Figure 4(f) shows the photocurrent measurements made upon visible light illumination using an AM1.5 solar simulator (1 sun) with a 450 nm longpass filter. The AuNP patterned $TiO_2$ shows approximately 1.5 times increase in photocurrent compared to bare $TiO_2$, owing to the hot carriers generated in the AuNPs. This is in stark contrast with electrodes with the PDMS "adhesion layer", which prevents current flow and hence no photocurrent can be generated (overlapping yellow and blue lines). Although beyond the scope of the current work, the system can be improved further for future studies dealing more rigorously with hot electron generation for photochemical processes. For instance, post treatments such as high temperature annealing of AuNPs printed on $TiO_2$ can help crystallize the $TiO_2$ film for increased conductivity, reduce the size of AuNPs for enhanced hot electron injection efficiency, and also improve the contact between the NP and film.

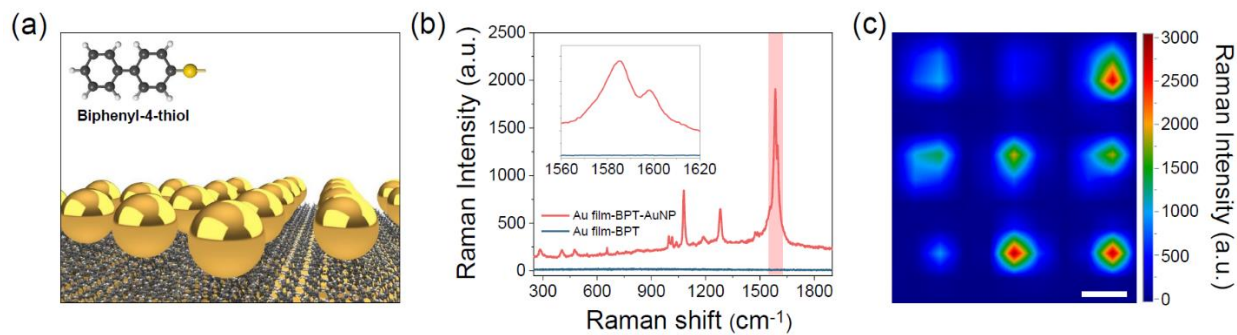

**Figure 5. Single particle SERS of BPT monolayer sandwiched between Au film and printed AuNP array.** (a) Schematic illustration of NPoM configuration with BPT monolayer. (b) SERS spectra of BPT monolayer from single AuNP on Au film. Inset shows the enlarged spectra for characteristic peaks of BPT. (c) 2D SERS mapping with high intensity signals from the vicinity of single AuNPs. Scale bar: 1 μm.

To broaden further the applicability of the TDIP technique, the AuNPs were printed on an Au film to create a nanoparticle-on-mirror (NPoM) configuration[55,56] for single particle SERS. The NPoM construct, illustrated in Figure 5(a), comprises a SERS-active biphenyl-4-thiol monolayer as an organic spacer between the AuNP and Au film mirror. This system has previously been proposed to give more reproducible signals than that comprising multimeric particles or antenna structures.[57,58] In these reports, the correlation between SERS enhancement and dark-field scattering was studied intensively by considering both the size and shape of AuNPs. However, existing NPoM constructs for SERS are based on drop casting the AuNPs which is prone to random deposition and thus requires navigation through the sample for analysis of single NPs. In contrast, patterning single AuNPs in a non-destructive manner can offer a more systematic and efficient way of sensing. Figure 5(b) shows the SERS spectra of the BPT probe molecule sandwiched between the printed AuNP and Au film, excited with a 633 nm laser. The presence of BPT is clear from the profile between 1580 cm$^{-1}$ and 1600 cm$^{-1}$ (enlarged in inset) showing two characteristic peaks. The split peaks pertain to the coupled vibrational mode of the two phenyl rings of the molecule.[57,58] The benefit of printing NPs non-destructively is that a Raman map of an ordered array can be obtained. This, for instance, would be challenging to achieve by other patterning methods such as optical printing, since the focused laser can potentially damage the molecules sitting on the Au film substrate and also deform the shape of the NP. As a simple demonstration to show that the BPT monolayer gives enhanced signals selectively at where the NPs are positioned to form the gap with the underlying Au film, a 2D Raman map was obtained for an array of printed AuNPs in Figure 5(c). The signals visualized correspond to the Raman shift range (inset of Figure 5(b)) that includes the two split peaks of BPT. As provided in Figure S6, no signal was obtained from the areas without the NPs, in par with previous reports.[57,58] From the 2D maps, the variation in Raman intensities arises due to inhomogeneity in the nanoparticles in terms of size and shape, as atomic defects and facets play a crucial role in SERS enhancements.[57,59] Given that achieving uniform signal intensity distribution yet remains a challenge using spheres due to their faceted nature,[57] patterning of nanocubes can be a future approach for obtaining a more homogeneous SERS enhancement[60] for such NPoM constructs. Overall, our results show that current approach to pattern arrays of NPs to create NPoM geometries has the potential to create molecular sensing platforms for the detection of probe molecules at the single particle level.

**CONCLUSION**

The integration of colloidal NPs in a deterministic and spatially accurate manner holds great promise in the creation of functional nanodevices for applications involving electrical circuits, molecular sensing, and photoelectrochemical processes for energy conversion, among others. In this work, TDIP was developed to pattern free-standing single AuNPs on solid surfaces with a high printing yield and accuracy onto various dielectric and conductive substrates without the use of any adhesion layer. Given that TDIP in itself shows no limit to the variety of NPs that can be assembled, the template dissolution-based printing method allows further flexibility with respect to the receiving substrate. This elevated flexibility has clear advantages over other positioning methods involving colloidal NPs. With the NPs in free-standing form, the capacity to have direct access to the NP/substrate interface allows the tuning of the plasmon resonance and the creation of electrical junctions for (photo)electrochemical processes. The technique is also non-destructive of the interface, in comparison to those using high power lasers or high voltages, enabling the configuration of NPoM geometry for detection of probe molecules residing at the gap between the printed NP and an Au film mirror. We envision that TDIP can serve as a valuable tool to create an extensive variety of interesting functional heterostructures and devices based on nanostructured colloids.

# METHODS

## *Materials*

Colloidal suspension of 100 nm diameter AuNPs was purchased from BBI solutions (UK). Poly(methyl methacrylate) (PMMA) with a nominal molecular weight of 100 kDa, used for creating the replica template from the Si master, was purchased from Polysciences, Inc. (Warrington, PA, USA). Potassium hexacyanoferrate (II) trihydrate and potassium hexacyanoferrate (III) used for electro-/photoelectro-chemistry and biphenyl-4-thiol used for single particle SERS were all purchased from Sigma-Aldrich (Merck KGaA, Darmstadt, Germany).

## *Fabrication of Si master template*

The silicon master template, consisting of ordered arrays of nanopillars, was fabricated by electron beam lithography (EBL). First, a freshly cleaned Si substrate was spin coated with a positive-tone resist (PMMA 950 A4, MicroChem, USA) of ~300 nm thickness and baked at 180 ºC for 5 min. The desired patterns were defined by electron beam exposure at 20 keV (Raith e_LINE, Raith GmbH, Germany), followed by a MIBK:IPA=1:3 development process. A metallic sacrificial mask was created by thermal evaporation of 80 nm think Cr at 0.1 Å s$^{-1}$, followed by lift-off in acetone at room temperature. Next, a reactive ion etching (RIE) recipe with a combination of $SF_6$ and $CHF_3$ was used to etch into the Si substrate for ~120 nm. The Si nanopillars were finally obtained by wet etching of the entire sacrificial hard mask with ceric ammonium nitrate-based etchant (Merck KGaA, Darmstadt, Germany). For large-area patterning in the nanoelectrochemistry application, a larger Si master template (1 cm$^2$) was fabricated by KrF photolithography (ASML KrF Scanner PAS 5500/700D, ASML, Netherlands) followed by $HBr/Cl_2$-based reactive ion etching to achieve a 90 nm trench depth (TCP-9400DFM, Lam Research, CA, USA).

*PMMA replica preparation*

PMMA solution was first prepared by dissolving the PMMA beads completely in acetone to yield a 5 wt% solution. To create the PMMA replica template, the PMMA solution was spin coated on the Si master at 3000 rpm for 60 s. The polymer film was then stripped off the master template using polyimide Kapton tape.

*Capillary-assisted particle assembly (CAPA)*

CAPA was achieved through a home-built set-up. The PMMA replica attached polyimide tape was fixed below a glass slide on a hot-plate set typically at 35-38 °C above the dew point. The temperature setting for nanoparticles is generally higher than for micron-sized particles to ensure that a dense accumulation zone is formed, considering the increase in particle diffusivity.[28] A small volume (100 μL) of AuNP suspension was trapped between the PMMA template and a clean glass cover slip. The cover slip was drawn across the surface of the tape using a piezoelectric motor at a speed of 1 μm s$^{-1}$. This formed a meniscus at the edge of the cover slip where the nanoparticles experienced the downwards and lateral forces necessary to immobilize nanoparticles into traps in the PMMA replica. Faster assembly speeds often resulted in low percentage of trapping and slower speeds led to some cases of random aggregates to form beyond the desired trapping sites. Throughout the assembly, the receding contact angle was monitored using a camera. The contact angle for the highest assembly yields was in the range of 45-47°.

### Printing of AuNPs on receiver substrates

The AuNP patterned assemblies obtained *via* CAPA were printed by manually pressing down the polyimide tape against pre-cleaned receiver substrates. The substrates were then fully immersed in an acetone bath for 1-1.5 h, after which the detachment of the polyimide tape from the substrate was observed as a result of the PMMA dissolving in acetone to leave behind NPs free-standing on the substrate.

### Optical characterization

SEM images were taken using a Hitachi S-4800 field emission scanning electron microscope. Dark-field images and scattering spectra were taken using a WITec confocal microscope (Germany) using a 100X air objective with 0.9 NA (Zeiss, Germany). Bright-field images were obtained using Olympus BX51M with a 0.9 NA 100X air objective (Olympus, Japan).

### Simulations for DF scattering

Finite-difference time-domain (FDTD) simulations using commercial Lumerical FDTD Solutions were performed for comparison to experimental results. 100 nm AuNPs positioned above Au, ITO and GaP films on Si or glass were illuminated with *s*- and *p*-polarized light, and the resulting data was combined to simulate unpolarized light. A total-field-scattered-field source was set at an angle of 70 degrees to normal to the substrate to imitate dark field illumination. The simulation was performed with single frequency illumination. Transmission through a 2D monitor behind the source collected transmission to the far field, integrated over a cone of interest with half angle 60.14° to simulate the NA=0.9, which was then normalized to the source power. A 1 nm gap size was used between the NP and film in all cases to simulate surface roughness. PML layers were incorporated to avoid reflection at boundaries and symmetric and antisymmetric boundary conditions were utilized to reduce computation time where symmetry conditions allowed. Built in optical properties of Au and $SiO_2$ were utilized (Johnson & Christy), numerical data was utilized for ITO[61] and ellipsometry data was used for GaP film simulations.[62]

*Electro- and photoelectrochemical analysis of ferro-/ferricyanide redox couple*

All electro- and photoelectrochemical measurements were conducted using an Ivium-n-Stat multichannel electrochemical analyzer (Ivium Technologies, Netherlands) in a three-electrode setup with Pt and Ag/AgCl as counter and reference electrodes, respectively. The reaction buffer was 0.1 M $KH_2PO_4$ (pH 7) containing 1 mM $K_3Fe(CN)_6$/$K_4Fe(CN)_6$. For cyclic voltammetry, the scan rate was set to 50 mV $s^{-1}$ for an applied potential window of –0.7 V to 1.1 V (*vs.* Ag/AgCl) to clearly visualize the redox peaks. For photocurrent measurements, a HAL-320 solar simulator was used (Asahi Spectra, Japan) along with a 450 nm longpass filter (Edmund Optics, USA). The light on and off intervals were set to 60 s.

*Single particle SERS measurements*

To form the biphenyl-4-thiol monolayer for SERS application, an Au film substrate was immersed in an ethanolic solution of BPT for 22 h. To remove any unbound molecules, the substrate was repeatedly rinsed and sonicated with ethanol and finally dried using $N_2$ gas. AuNP assemblies were then printed using the provided methodology. Single particle SERS was conducted using a LabRAM HR Evolution microscope for high-resolution Raman (Horiba, Japan) equipped with a 633 nm laser. The laser power and integration time were set to 0.31 mW and 10 s, respectively.

## ACKNOWLEDGMENTS

J.L. and Y.N. acknowledge support from the Basic Science Research Program through the National Research Foundation of Korea (NRF) funded by the Ministry of Science, ICT & Future Planning (NRF-2020R1A2C2004168). A.R. acknowledges the financial support of the Royal Society (UF150542 and IEC/R2/170185). S.A.M. and E.C. acknowledge funding and support from the Deutsche Forschungsgemeinschaft (DFG, German Research Foundation) under Germany´s Excellence Strategy – EXC 2089/1 – 390776260, the Bavarian program Solar Energies Go Hybrid (SolTech), the Center for NanoScience (CeNS) and the European Commission through the ERC Starting Grant CATALIGHT (802989). The authors also acknowledge support from EPSRC - UK under the grant (EP/ M013812/1, Reactive Plasmonics). S.A.M. additionally acknowledges the Lee-Lucas Chair in Physics.

## SUPPORTING INFORMATION AVAILABLE

Analysis of assembly and printing yields, statistical analysis of printing accuracy, images of printed nanocubes, scheme and example from alternative fabrication approach (with supplementary note), simulated dark-field scattering with separate *s*- and *p*-polarized illumination for single AuNP on GaP film, and single particle SERS for an array with a missing particle are provided in the Supporting Information. This material is available free of charge *via* the Internet at http://pubs.acs.org.

**TOC FIGURE:**

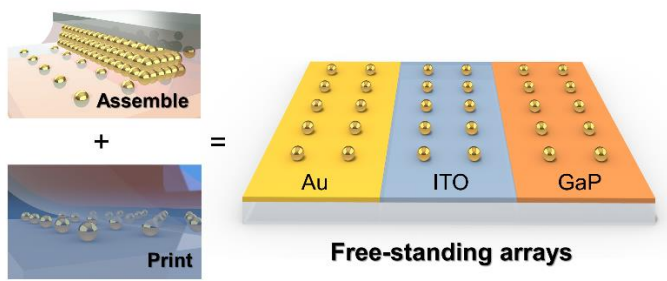